\documentclass[conference]{IEEEtran}
\IEEEoverridecommandlockouts
\usepackage{cite}
\usepackage{amsmath,amssymb,amsfonts}
\usepackage{algorithmic}
\usepackage{graphicx}
\usepackage{textcomp}
\usepackage{xcolor}
\usepackage{amsmath,amssymb,nicefrac,braket,graphicx,subcaption,xfrac,dirtytalk}
\def\BibTeX{{\rm B\kern-.05em{\sc i\kern-.025em b}\kern-.08em
    T\kern-.1667em\lower.7ex\hbox{E}\kern-.125emX}}
\begin{document}

\title{Approximate Phase Search and Eigen-Estimation using Modified Grover's Algorithm}

\author{\IEEEauthorblockN{Sayantan Pramanik, M Girish Chandra, Shampa Sarkar and Manoj Nambiar}
	\IEEEauthorblockA{\textit{TCS Research and Innovation,}
		\textit{India}\\
		\{sayantan.pramanik, m.gchandra, shampa.sarkar, m.nambiar\}@tcs.com}
}

\maketitle

\begin{abstract}
An attempt has been made in this paper to modify Grover’s Algorithm to find the binary
string solutions approximating a target cost value. In that direction, new
Controlled Oracle and the Local Diffusion Operator are suggested, apart from
incorporating suitable ancilla qubits. A possible strategy to estimate eigenvalues
and eigenstates of a given cost Hamiltonian, extending the reasoning of the
methodology, is also pointed out. Typical results and relevant discussions are
captured to support the propositions.
\end{abstract}
\medspace
\begin{IEEEkeywords}
Grover's algorithm, aproximate search, phase oracle, global diffusion operator, local diffusion
\end{IEEEkeywords}

\section{Introduction}\label{sec:intro}
Grover’s Algorithm was one of the flagship algorithms that demonstrated a
quantum advantage over its classical counterparts. Right from its proposition
more than two decades ago, it has attracted undiminished interest and
explorations from the Quantum Information Processing research community \cite{g1, phasematching, g2, weightedtargets, gas1, gas2, ibmjpmc, partialdiffusion, counting, jpmc}.
Many interesting extensions and modifications resulted out of this focus. In this
paper, we considered the possibility of working out different (all) possible
solutions of binary strings approximating the target cost of a function to varied
degrees, within the context of combinatorial optimization setting. The idea is to
pick up the solutions which satisfy the cost more closely with greater probability
than others. In this direction, we propose a simple extension in terms of
nonbinary phase oracles that append phases to the states based on the cost of
the state, compared to the traditional binary marking in the original Grover’s
Algorithm. Two more tricks appeared to be necessary to solve the problem-
expansion of the search space using suitable number of ancilla qubits and a
novel definition of what we refer as Local Diffusion Operator. As in other Grover-
based algorithms, applying suggested nonbinary phase oracle (which is a
controlled oracle) and the Diffusion operators successively, we obtained the
useful amplitude amplification of the approximate states by measuring the
requisite qubits in the computational bases, leaving out the ancillae. Another spin
off this extension is to further parametrize the Control Oracle to estimate the
eigenvalues and the corresponding eigenstates of the cost Hamiltonian.

The paper is organized as follows: In Section \ref{sec:grover}, necessary points on Grover’s
Algorithm required as a base for our work are captured. Section \ref{sec:aps} discusses at
length the proposed method, including Control Oracle and Local Diffusion
Operator. Some observations on the number of iterations required is also brought
out. Section \ref{sec:eigen} contains details about the eigen-estimation. The results achieved
so far, relevant remarks and possible further examinations are discussed in
Section \ref{sec:results}.

\section{Grover's Algorithm}\label{sec:grover}
Grover's Algorithm \cite{grover} addresses the problem of finding the \textit{needle in a haystack}, i.e., searching for an element in an unstructured database. It provides a quadratic speedup over classical search algorithms, with a complexity of $O(\sqrt{N})$, where $N$ is the size of the database.

Given a database with $N$ entries, a set $\Omega$ of $k$ target states denoted by $\omega_j$, and an \textit{oracle} that has the capability to identify and mark them, the algorithm finds and returns the marked elements with near-certain probability, with approximately $\sqrt{\frac{N}{k}}$ calls to the oracle.

Steps to follow in Grover's algorithm:

\begin{enumerate}
	\item Application of the Hadamard gate to place all the qubits in an equal superposition state $\ket{s}$. 
	\item Applying the oracle $O_G$ on the qubits:
	\begin{equation}
		O_G = I - (1 - e^{-i\alpha}) \sum_{j=1}^{k}\ket{\omega_j}\bra{\omega_j}
	\end{equation}
	\item Applying the diffusion operator $D$ to selectively increase the probability of obtaining the marked states through amplitude amplification:
	\begin{equation}
		D = (1 - e^{-i\beta}) \ket{s}\bra{s} - I
	\end{equation}
	\item Performing $r$ repetitions of steps $2$ and $3$, where $r$ is a function of $\alpha$, $\beta$, $k$, and $N$, and is in the order of $\sqrt{\frac{N}{k}}$.
\end{enumerate}

It was identified by Long, et al. in \cite{phasematching}, that the search algorithm is feasible only when $\alpha = \beta$, and the condition is known as phase-matching.

\section{Approximate Phase Search}
\label{sec:aps}
In this paper, we endeavour to solve the search problem spelt out in Section \ref{sec:intro}, which we also refer to as Approximate Phase Search using a modified version of Grover's search algorithm. In literature there exists a technique named \textit{fuzzy string searching}, also called the Approximate String Matching Algorithm (ASMA). As the name suggests, ASMA tries to find strings that are approximately related to a given string, which are found through insertion, deletion, substitution or transposition of certain characters within the original string. For instance, the words \textit{paste}, \textit{past}, \textit{part}, \textit{pass}, etc., can all be derived from the word \textit{past} through the use of said operations.

Similar to its string counterpart, the approximate phase search algorithm attempts to solve an equivalent problem in the combinatorial regime. Given a set of bitstrings $X = \{x_j\}$, a function $c(x_j)$ to calculate the \textit{cost} of each bitstring, and a target cost $c_0$, the goal is to find the bitstring with cost closest to $c_0$ with the highest probability. More generally, the probability of measuring $x_j$ after the algorithm, $P(x_j)$ should be inversely proportional to $|c(x_j)-c_0|$. 

In \cite{weightedtargets}, the authors talk about a variation of Grover's algorithm where the target states are \textit{weighted}. The method  was devised so that the \say{\textit{probability of getting each target can be approximated to the corresponding weight coefficient $w_j$}}. With $\ket{q} = \sum b_j \ket{\omega_j}$ being the weighted superposition of all the states, their oracle was defined as $O_W = I - 2 \ket{q}\bra{q}$, where:

\begin{equation}
	b_j =
	\begin{cases}
		\sqrt{w_j},& \text{if } \omega_j \in \Omega\\
		\ \ \ 0,              & \text{otherwise}
	\end{cases}
\end{equation}

However, there appear to be two major flaws in this approach. The first being that the oracle should have been $O_W = I-2\sum |b_j|^2 \ket{\omega_j}\bra{\omega_j}$. Even with the corrected oracle, it is not unitary if $b_j$ is not a complex number with unit modulus.

Finally, it is worth mentioning that approximate phase search is principally different from the Grover-based optimization algorithms discussed in \cite{gas1,gas2,ibmjpmc}. The optimization algorithms aim at obtaining the target states (having minimum cost, or equivalently, having cost $c_0$) with as high a probability as possible, while effectively reducing the probability amplitudes of all the other states to zero. If a state with cost $c_0$ does not exist, then the optimization algorithms tend to fail. On the contrary, all the states in approximate phase search are potential solutions to the problem with varying measures of accuracy which are defined by their absolute deviation from the target cost. Higher the deviation, lower should be the probability of measuring the corresponding state. As a result, even if a state with the target cost does not exist, the closest possible match may be obtained from the algorithm.

\subsection{Non-Binary Phase Oracles}

Grover's algorithm uses an oracle $O_G$ that identifies and marks the elements being searched for with a negative sign. In the phase-matched version of the algorithm, the corresponding states are appended with a phase of $\alpha$. Essentially, a binary-phase oracle is used which has the following effect:

\begin{equation}
	O_G(x)\ket{x} = e^{i\Pi(x)}\ket{x}
\end{equation}
where $\Pi(x) = 0$ when $x \notin \Omega$, or $\Pi(x) = \alpha$ when $x \in \Omega$. When $\alpha = \pi$, we recover Grover's original algorithm.

In the case of Approximate Search, we propose the use of \textit{non-binary} phase oracles that append phases to the states based on their \textit{cost}:

\begin{equation}
	O\ket{x} = e^{i\Phi(x)}\ket{x}
\end{equation}
where $\Phi(x) = f(c(x))$ is a function which is designed in such a way that it depends on the cost of the state $x$, and is scaled appropriately to append the phase $\pi$ to the best solution states.

\subsection{Expansion of Search Space}
\label{sec:expansion}
Grover's algorithm acts by amplifying the amplitude of the states with phase $\pi$. The amplification works by reducing the amplitudes of the non-targeted states, which reduces the probability of measuring them. In the context of  search, most of the states are expected to have a non-zero phase and are potential solutions to the problem. This makes the redistribution of amplitude a formidably difficult task to accomplish. The difficulty can be overcome by expanding the search space through the use of ancilla qubits, which is unintuitive from the perspective of a search problem, but as demonstrated later, does not increase the order of complexity of the problem.

The ancillae are appended to the system and are ignored at the time of measurement, which acts as a form of marginalisation. In the light of the expanded search space, the oracle is converted into a controlled gate which acts on the \textit{work qubits} only when all the ancillae are in the state $\ket{1}$. The action of the controlled oracle $O_C$ on a state $\ket{x}$ in a system with $m$ ancilla qubits, which are in the uniform superposition state $\ket{+}_m$, can be represented as:

\begin{equation}
	O_C\ket{x}\ket{+}_m = \sum_{j=0}^{2^m-2}\ket{x}\ket{j} + e^{i\Phi(x)}\ket{x}\ket{1}_m
\end{equation}

\subsection{Local Diffusion Operator}
\label{sec:local}
The diffusion operator $D$, defined by Grover, acts on all of the work-qubits to perform \textit{inversion about the mean}. It has a \textit{global view of the system} in the sense that it acts on all the states at the same time to perform the reflection operation. In a similar manner, we define a local diffusion operator $D_L$ that has a \textit{local view} of the system. If there are $m$ ancilla qubits, then the local diffusion operator acts on $2^m-1$ states that do not have a phase, and only one state with a phase associated with it. Thus, it works only in the locality of the superposition states of the ancilla qubits for a given state of the work qubits. The local diffusion operator being proposed here is not the same as the partial diffusion operator discussed in \cite{partialdiffusion}.

The local diffusion operator is a block-diagonal, $2^{m+n} \times 2^{m+n}$ unitary matrix with $2^m \times 2^m$ dimensional diffusion operators forming its diagonal elements, when the ancilla qubits are present at the end of the system:

\begin{equation}
\label{eq:localDiffusion}
\begin{split}
D_L &= 2\sum_{j=0}^{2^n-1}\ket{j}\ket{+}_m\bra{+}_m\bra{j}-I	\\
&= 2\sum \ket{j}\ket{+}_m\bra{+}_m\bra{j} - \sum_{j=0}^{2^n-1} \sum_{j'=0}^{2^m-1} \ket{j}  \ket{j'} \bra{j'} \bra{j}	\\
&= 2\sum \ket{j} \bra{j} \otimes \ket{+}_m \bra{+}_m - \sum \ket{j} \bra{j} \otimes \sum \ket{j'} \bra{j'}\\
&= \sum_{j=0}^{2^n-1}\ket{j}\bra{j} \otimes (2\ket{+}_m \bra{+}_m - I)\\
&= I \otimes (2\ket{+}_m \bra{+}_m - I)
\end{split}
\end{equation}

From Equation \eqref{eq:localDiffusion} it is clear that the local diffusion operator is analogous to its global counterpart working on just the ancilla qubits on the system. Since the local diffuser acts on groups of $2^m$ states, it is imperative to have at least two ancillae, as it is known that performing diffusion on only one qubit does not provide any measurable advantage in the computational basis. Application of operators $O_C$ and $D_L$ in succession $\lfloor \frac{\pi}{4} \sqrt{2^m} \rfloor$ times acts as a \textit{preprocessing} step. It increases the amplitude of the states with phase $\pi$ (i.e., the solution states) from $\sfrac{1}{\sqrt{2^{m+n}}}$ (in the equal superposition state at the start of the algorithm) to nearly $\sfrac{1}{\sqrt{2^n}}$. Parallely, it also amplifies the states in which the ancilla qubits are in $\ket{1}_m$ at the cost of others. 

\subsection{Overview of the Approximate Phase Search Algorithm}

Having discussed about all the components required by the approximate phase search algorithm, it might now be ideal to list down the requisite steps for a complete overview of the algorithm:

\begin{enumerate}
	\item Initialization of a quantum circuit with two quantum registers, one with $n$ work qubits and the other with $m$ ancilla qubits, along with a classical register containing $n$ classical bits.
	\item Bringing the system to equal superposition through the use of $H$ gates on all of the qubits.
	\item Repetition of the following steps $\lfloor \frac{\pi}{4} \sqrt{2^m} \rfloor$ times, which carries out the requisite preprocessing:
	\begin{enumerate}
		\item Application of the controlled-oracle with the ancillae as control and the work qubits as the target.
		\item Followed by the application of the local diffusion operator.
	\end{enumerate}
	\item Looping over the following two steps $N_{iter}$ times:
	\begin{enumerate}
		\item Again, applying the oracle as discussed in the previous point.
		\item application of the global diffusion operator, this time, on all the qubits in the system.
	\end{enumerate}
	\item Measurement of the $n$ work qubits into the $n$ classical bits.
\end{enumerate}

The ideal number of ancillae, $m$, and iterations, $N_{iter}$, required for the optimal performance of the algorithm is subject to further research. As there are $2^n$ possible phases appended by the oracle, as opposed to just two in Grover's traditional algorithm, that interfere with each other, it is suspected that the ideal number of iterations may not have a closed-form solution. Additionally, the degeneracy of the system need to be considered to find $N_{iter}$. To that extent, a generalized version of the Quantum Counting Algorithm \cite{counting} for multiple phases needs to be designed, and is beyond the scope of this work. To simplfy the application and to demonstrate the feasibility of the algorithm, we emperically set $m=n$, and $N_{iter}=\sqrt{2^n}$, which yields consumable, if not the best, results.

\begin{figure*}
	\centering
	\includegraphics[scale=0.315]{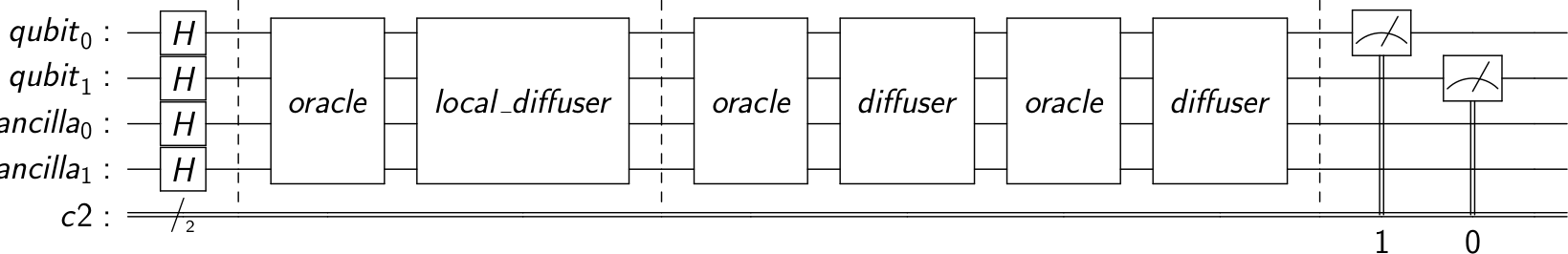}
	\caption{Quantum Circuit used to perform Approximate Phase Search for 2 work, and 2 ancilla qubits.}
	\label{fig:circuit}
\end{figure*}

The circuit for the complete scheme for $n=m=2$ qubits can be found in Figure \ref{fig:circuit}.

\subsection{Kullback-Leibler Divergence as a Performance Metric}
\label{sec:KLD}
Kullback-Leibler Divergence (in short, KLD) is a metric used to compare two probability distributions, i.e., the extent to which the given distributions differ from eachother. Its value ranges between $0$ and $\infty$, $0$ when the two distributions are exactly identical. Given two distributions, $P(x)$ and $Q(x)$, the KLD between them is specified by:

\begin{equation}
	D^{KL} = \sum_j P(x_j) \log(\frac{P(x_j)}{Q(x_j)})
\end{equation}

After running Grover's algorithm for multiple shots, the final, normalised histogram of measurements can be treated as a probability distribution, $P(x)$. The other distribution in question, $Q(x)$, is the histogram obtained from an equal superposition of all the the states, and is basically an uniform distribution. The reason behind selecting the uniform distribution to compare $P(x)$ against is that the performance of any variant of Grover's algorithm relies on the ability of the oracle to properly mark the target states (with various phases, in our case). Failing to appropriately mark the states will result in none of them getting selectively amplified, yielding the uniform distribution as a result. This allows us to use $D^{KL}$ as a measure of success of the algorithm, provided that the state acieved with highest probability is one of the targeted states.

It must also be noted that the KLD metric reduces with the introduction of degeneracy into the system. As a general rule, higher the degeneracy, lower is the value of KLD.

\section{Eigen-estimation using the proposed Approximate Phase Search}
\label{sec:eigen}
Given a cost Hamiltonian, $H_C$, we attempt to find its eigenvalues and corresponding eigenvectors (for ground and excited states) by leveraging the approximate phase search algorithm that has been described thus far. To accomplish that, a \textit{parameterized} oracle with parameter $\lambda$, is defined as:

\begin{equation}
	\label{eq:oracleHc}
	O_{H_C} = e^{i \pi \sfrac{H_C}{\lambda}}
\end{equation}

If the Hamiltonian $H_C$ is Hermitian, then the oracle $O_{H_C}$ is guaranteed to be unitary for real values of $\lambda$. The oracle can be converted to a controlled one as discussed in subsection \ref{sec:expansion}. If the eigenvalues and eigenvectors of $H_C$ are given by $\lambda_j$ and $\ket{\psi_j}$, then $O_{H_C}$ can be written in the form:

\begin{equation}
	O_{H_C} = \sum_j e^{i \pi \sfrac{\lambda_j}{\lambda}} \ket{\psi_j} \bra{\psi_j}
\end{equation}
For $\lambda = \lambda_j$, we find that the oracle appends a phase of $\pi$ to the corresponding eigenstate $\ket{\psi_j}$. If the search algorithm is run for some arbitrary $\lambda$, and $D^{KL}_\lambda$ is the resultant KLD (as discussed in subsection \ref{sec:KLD}), then we hypothesize that for the appropriate number of iterations,

\begin{equation}
	D^{KL}_{\lambda_j} > D^{KL}_{\lambda_j\pm \epsilon}
\end{equation}
where $\epsilon$ is a small positive change in the value of $\lambda$.

As a consequence of running the search algorithm for every value of $\lambda$ with a suitable step-size within a specified range, on a plot of $D^{KL}_{\lambda}$ against $\lambda$, we expect to get peaks for $\lambda = \lambda_j$, and in the case of each peak, the state $\ket{\psi_j}$ is obtained with the highest probability. Thus, within a range, all of the eigenvalues and eigenvectors of the system can be obtained by looping over a single parameter, without having to find the ground state first and then finding each subsequent excited state using variational algorithms \cite{vqe,qaoa} as discussed in \cite{excited}.

\section{Implementation and Results}
\label{sec:results}
To implement and test the algorithms presented in this paper, we have made use of Qiskit, an open-source framwork for programming quantum computers, by IBM \cite{qiskit}. Specifically, their noiseless \textit{qasm\_simulator} backend has been utilised for running the quantum circuits with a high number of \textit{shots}. We present implementations and results for two \textit{NP-complete} problems: a variation of the subset-sum problem, and the max-cut problem.

\subsection{Construction of Local Diffuser}

As discussed in section \ref{sec:local} and as evident from Equation \eqref{eq:localDiffusion}, the local diffusion operation, on a system of $n$ work qubits and $m$ ancillae  at the end, is equivalent to applying Grover's diffusion on the ancilla qubits alone.

\begin{figure}
	\centering
	\includegraphics[scale=0.4]{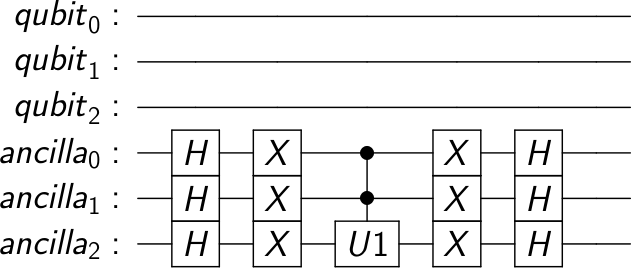}
	\caption{Local Diffusion Operation on a circuit with $m=n=3$ qubits}
	\label{fig:local}
\end{figure}

Figure \ref{fig:local} shows the gate-circuit implementation for a simple system with six qubits, three of them used to enode the problem, and the other three to increase the dimension of the search space, formulated in Section \ref{sec:expansion}. The $H$ and $X$ gates can be merged for a more efficient circuit with lesser depth, as suggested in \cite{jpmc}. Also, the $U1$ gate has a parameter of $\pi$.

\subsection{A Variation of Subset-Sum Problem}
\label{sec:subsetsum}
Given a set of $n$ real numbers, $\xi = \{s_0, s_2, \dots, s_{n-1}\}$, the problem is to find $\zeta \subseteq \xi$ such that the sum of elements in $\zeta$ is as close as possible to a given real number $S$. Trivially, $2^n$ subsets of $\xi$, can be represented by the binary strings $\{0,1\}^n$, where each bitstring $x_j = y_0 \ y_1 \dots y_{n-1}$; $y_l =1$ if the element $s_l$ is present in the $j^{th}$ subset, and $y_l = 0$ otherwise. The cost $c(x_j)$ of the bitstring $x_j$ is given by $c(x_j) = \sum_{l=0}^{n-1} s_l y_l$.

\begin{figure}
	\centering
	\begin{subfigure}[b]{0.49\textwidth}  
		\centering 
		\includegraphics[scale=0.6]{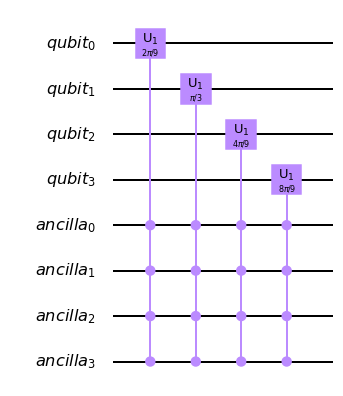}
		\caption{Oracle}
		\label{fig:subsetsumora}
	\end{subfigure}
	\begin{subfigure}[b]{0.48\textwidth}  
		\centering 
		\includegraphics[scale=0.51]{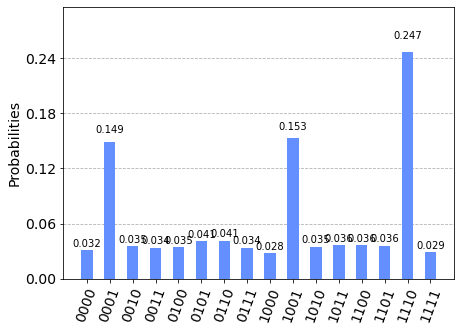}
		\caption{Histogram of Results}
		\label{fig:subsetsumhist}
	\end{subfigure}
	\caption{Oracle and results for the example problem discussed in Section \ref{sec:subsetsum}}
	\label{fig:results}
\end{figure}

To construct the oracle for this problem, the function $\Phi(x_j)$ in Equation \eqref{eq:subsetphi} is chosen such that it linearly scales the cost to map $c(x_j) = S$ to the phase $\pi$. It has the disadvantage that if $c(x_j)$ is an odd multiple of $S$, the effective phase evaluates to $\pi$. As a simple example, let $\xi = \{2,3,4,8\}$ and $S=9$. The oracle and results have been portrayed in Figures \ref{fig:subsetsumora} and \ref{fig:subsetsumhist}, respectively. It can be seen that state $\ket{1110}$ with the highest probability has a cost of $9$, and the second and third most probable states have costs of $10$ and $8$, each. It is evident that our new algorithm is able to \textit{search} the approximate solutions.

\begin{equation}
	\label{eq:subsetphi}
	\Phi(x_j) = \pi \frac{c(x_j)}{S}
\end{equation}

\subsection{Max-cut Problem}
\label{sec:maxcut}

\begin{figure}
	\centering
	\begin{subfigure}[b]{0.49\textwidth}  
		\centering 
		\includegraphics[scale=0.5]{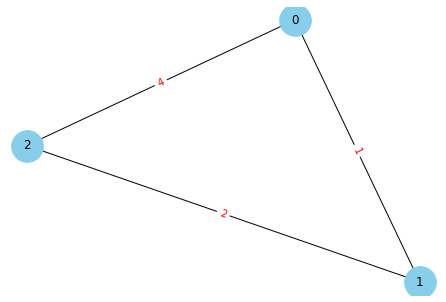}
		\caption{Undirected Graph}
		\label{fig:graph}
	\end{subfigure}
	\begin{subfigure}[b]{0.49\textwidth}  
		\hspace{-0.6cm}
		\includegraphics[scale=0.4]{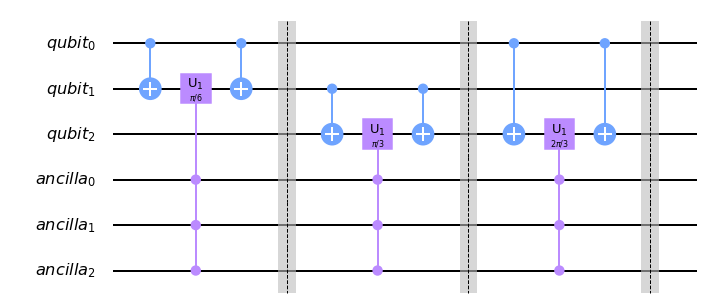}
		\caption{Oracle for the graph in subfigure a}
		\label{fig:maxcutoracle}
	\end{subfigure}
	\caption{Graph and corresponding oracle for the example problem discussed in Section \ref{sec:maxcut}}
	\label{fig:maxcutfigures}
\end{figure}

\begin{figure}
	\centering
	\includegraphics[scale=0.45]{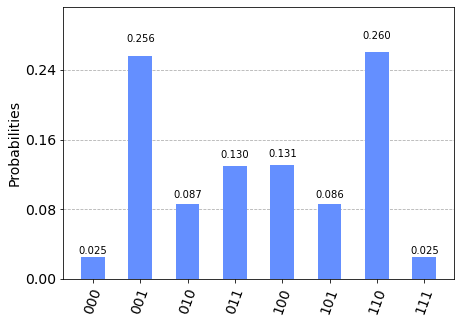}
	\caption{Histogram of max-cut results for the graph in \ref{fig:graph}}
	\label{fig:maxcutresults}
\end{figure}

The max-cut problem starts with an undirected graph $G(V,E)$ with a set of vertices $V$ and a set of edges $E$ between the vertices. The weight $w_{jl}$ of an edge between vertices $j$ and $l$, is a positive real number, with $w_{jl} = 0$ if there is no edge between them. A cut is a set of edges that separates the vertices $V$ into two disjoint sets $V_{1}$ and $V_{2}$, such that $V_{1} \subseteq V$ and $V_{2}=V \setminus V_{1}$, and the cost of a cut is defined as the as the sum of all weights of edges connecting vertices in $V_{1}$ with vertices in $V_{2}$. One define a cut for the graph by labelling the vertices with $s_{j}$ such that $s_{j} = 1$ suggesting that node $j$ belongs to $V_{1}$ and $s_{j} = 0$ corresponds to $V_{2}$ (of course, $V_{1}$ and $V_{2}$ can be interchanged). The cost Hamiltonian for the max-cut problem is specified as \cite{maxcut}:

\begin{equation}
	H_C=\sum_{jl}\frac{w_{jl}}{2}(I - \sigma^z_j \sigma^z_l)
	\label{eq:maxcutCost}
\end{equation}
The aim is to partition the nodes in such a way that the cost of the resulting cut is maximized. The cost Hamltonian can be used to construct the oracle by leveraging Equation \eqref{eq:oracleHc}. This step is tricky for problems like max-cut where the cost of the best cut is not known beforehand. If the max-cost is known and is equal to $C_{max}$, then $\lambda$ is set equal to $C_{max}$. Otherwise, the strategy to circumvent this barrier is to start with the observation that bipartite graphs have the highest possible max-cut cost, which is equal to the sum of all the weights of the edges in the graph. Then setting $\lambda = \sum_{jl} w_{jl}$, there is a hope that the state with highest cost is amplified to a greater extent towards higher probability of occurance. Of course, one can expect the deterioration of performance with increasing difference between $\lambda$ and the highest cost.
For max-cut, too, the graph structure, oracle, and results for $\lambda=6$ are demonstrated for a toy problem (due to size constraints on the quantum circuit) in Figures \ref{fig:graph}, \ref{fig:maxcutoracle} and \ref{fig:maxcutresults}, respectively. The results do show the possibility of identifying the correct states in this toy example.

\subsection{Finding Eigenvalues}

\begin{figure}
	\centering
	\includegraphics[scale=0.65]{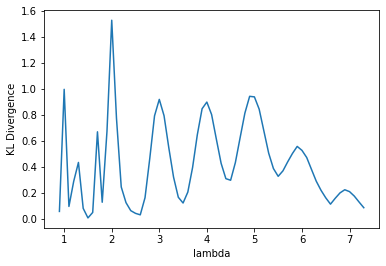}
	\caption{Plot of KLD against $\lambda$, where peaks can be seen when the value of $\lambda$ becomes equal to an eigenvalue of the system.}
	\label{fig:eigen}
\end{figure}

For the eigen-estimation, we return to the subset-sum problem where we attempted to find the eigenvalues for the corresponding problem Hamiltonian in Equation \eqref{eq:subsetHc}:

\begin{equation}
	\label{eq:subsetHc}
	H_C = \sum_j s_jZ_j
\end{equation}
where $s_j$ is the $j^{th}$ element in $\xi$, $Z = \sfrac{(I - \sigma^z)}{2}$ and $Z_j$ is the appropriate tensor product of $I$ and $Z$. A set $\xi = \{1,1,1,1,1,1,1\}$ was considered, and it is trivial to see that the set of eigenvalues are $\Lambda = \{0,1,2,3,4,5,6,7\}$, with each eigenvalue $\lambda_j$ having a degeneracy of $k_j = $ $7 \choose \lambda_j$. The plot for Kullback-Leibler Divergence against $\lambda$ has been shown in Figure \ref{fig:eigen}, which was obtained on looping over values of $\lambda$ with a small step-size $\epsilon$. The number of global iterations were  set to $\sqrt{\frac{N}{k_j}} \pm 1$ and the best values were captured in the plot. Additional measures had to be taken to curtail the effects of higher harmonics, as discussed in Section \ref{sec:subsetsum}. Some anomalies still appear in the plot, such as the abnormally high peak for $\lambda_j=2$, and the decaying values for $\lambda_j=6$ and $\lambda_j=7$. Finally, it needs to be mentioned that comparably good results for eigen-estimation were obtained when all the elements in the set $\xi$ were equal. With arbitrary values, the effect of interference dominated the plot and the algorithm needs to be investigated further for improvements.

\section{Conclusion}
An effort has been made in this paper to arrive at suitable modifications to the standard Grover’s Algorithm to get different possible binary strings as solutions for approximating the target cost of a function within the context of combinatorial optimization; they approximate the cost to different extents. The relevant results of the proposal for subset-sum and max-cut demonstrate the applicabilty and usefulness of the research. The methodology also facilitates the estimation of eigenvalues and eigenstates of the cost Hamiltonian, but needs additional examination towards obtaining good results in different scenarios. Further study is also underway to strengthen the present emperical suggestion for the number of iterations.

\section{Acknowledgement}
The authors sincerely thank Dr. Gautam Shroff, Head, TCS Research, for his encouragement and support.

\vspace{12pt}


\begin{thebibliography}{00}
\bibitem{grover} L.K. Grover, ``A fast quantum mechanical algorithm for database search,'' Proceedings, 28th Annual ACM Symposium on the Theory of Computing, (May 1996) p. 212.
\bibitem{g1} L. GuiLu, Z. WeiLin, L. YanSong and N. Li, “Arbitrary phase rotation of the marked state cannot be used for grover's quantum search algorithm,” 1999 Commun. Theor. Phys. 32 335.
\bibitem{phasematching} G.L. Long, Y.S. Li, W.L. Zhang, L. Niu, ``Phase matching in quantum searching,'' Physics Letters A, vol. 262, 1999.
\bibitem{g2} F.M. Toyama, W. van Dijk, Y. Nogami, M. Tabuchi, Y. Kimura, “Multi-phase matching in the grover algorithm,” Physical Review A, (Apr 2008), vol. 77.
\bibitem{weightedtargets} L. Panchi and L. Shiyong, "Grover quantum searching algorithm based on weighted targets," in Journal of Systems Engineering and Electronics, vol. 19, no. 2, pp. 363-369, April 2008, doi: 10.1016/S1004-4132(08)60093-6.
\bibitem{gas1} W. P. Baritompa, D. W. Bulger, and G. R. Wood, “Grover’s quantum algorithm applied to global optimiza- tion,” SIAM J. on Optimization 15, 1170–1184 (2005).
\bibitem{gas2} D. Bulger, W. P. Baritompa, and G. R. Wood, “Im- plementing pure adaptive search with grover’s quantum algorithm,” Journal of Optimization Theory and Appli- cations 116, 517–529 (2003).
\bibitem{ibmjpmc} A. Gilliam, S. Woerner, C. Gonciulea, ``Grover adaptive search for constrained polynomial binary optimization,'' arXiv:1912.04088 [quant-ph]
\bibitem{partialdiffusion} A. Younes, J. Rowe, and J. Miller, ``Quantum search algorithm with more reliable behaviour using partial diffusion,'' AIP Conference Proceedings 734, 171 (2004).
\bibitem{counting} G. Brassard, P. Hoyer, A. Tapp, ``Quantum Counting,'' Lecture Notes in Computer Science (1998), Springer Berlin Heidelberg, p. 820–831.
\bibitem{vqe} A. Peruzzo, J. McClean et al., “A variational eigenvalue solver on a photonic quantum processor”. Nature Communications 5, 4213 (2014).
\bibitem{qaoa} E. Farhi, J. Goldstone, S. Gutmann, “A quantum approximate optimization algorithm”. arXiv:1411.4028 [quant-ph].
\bibitem{excited} O. Higgott, D. Wang, S. Brierley, “Variational quantum computation of excited states”. Quantum, vol. 3, (July 2019), p. 156.
\bibitem{qiskit} H. Abraham et. al., “Qiskit: an open-source frame- work for quantum computing,” (2019).
\bibitem{jpmc} A. Gilliam, M. Pistoia, C. Gonciulea, “Optimizing quantum search using a generalized version of grover's algorithm,” arXiv 2005.06468 [quant-ph].
\bibitem{maxcut} L.Tse, P. Mountney, P. Klein, S. Severini, “Graph Cut Segmentation Methods Revisited with a Quantum Algorithm,” arXiv:1812.03050.
\end{thebibliography}
\end{document}